# Radio Astronomy Transformed: Aperture Arrays – Past, Present & Future

## M.A. Garrett[1]

*ASTRON, Netherlands Institute for Radio Astronomy*
*Postbus 2, 7990 AA, Dwingeloo, The Netherlands*
*E-mail:* `garrett@astron.nl`

I review the early development of Aperture Arrays and their role in radio astronomy. The demise of this technology at the end of the 1960's, and the reasons for the rise of parabolic dishes is also considered. For almost 40 years, Aperture Arrays languished as historical curiosities while large single dishes or arrays of dishes dominated the radio astronomy agenda. The parallels with the Antikythera mechanism (see these proceedings) are briefly presented. Aperture Arrays re-entered the world of radio astronomy as the idea to build a huge radio telescope with a collecting area of one square kilometre (the Square Kilometre Array, SKA) were debated within ASTRON and the rest of the radio community around the world. Huge advances in signal processing, digital electronics, high-speed networking and high-performance computing systems had transformed Aperture Arrays in terms of their capability, flexibility and reliability. In the mid-1990s, ASTRON started to develop and experiment with the first high frequency aperture array tiles for radio astronomy. Operating at frequencies of up to 1-2 GHz for studies of neutral hydrogen, tiled systems such as AAD, OSMA, THEA & EMBRACE were built and tested. In the slipstream of these efforts, Phased Array Feeds (PAFs) for radio astronomy were invented and LOFAR itself emerged as a next generation telescope and a major pathfinder for the SKA. Meanwhile, the same advantages that aperture arrays offered to radio astronomy (large multiple field-of-views, rapid electronic steering, reliability, durability, flexibility, cost and performance) had already made dishes obsolete in many different civilian and military applications. The first commissioning results from LOFAR and other Aperture Arrays (MWA, LWA and PAPER) currently demonstrate that this kind of technology can transform radio astronomy over 2 decades of the radio spectrum, and at frequencies up to at least 1.5 GHz. This "reinvention of radio astronomy" has important implications for the design and form of the full SKA. Building a SKA that is simply the "VLA on steroids" is simply not good enough – we have the ability to do much, much better. Like the Antikythera mechanism itself, we must amaze future generations of astronomers – they and the current generation deserve nothing less.

---

[1] Speaker

## 1. Early days: Aperture Arrays and radio astronomy

In its simplest form – an array of connected dipoles, Aperture Arrays are as old as radio astronomy itself. From the broadside array of Jansky, to the distributed interferometer that is LOFAR, these systems have made some of the most important discoveries in our field. Discoveries include the detection of radio emission from the Sun and Jupiter; the first lunar and planetary radar measurements, the first interferometer experiments, the detection of meteorite trails with velocities bounded by the solar system escape velocity, plus the first systematic surveys of the sky revealing new phenomena such as radio galaxies, quasars and pulsars, to name only a few (see [1] for the early history of radio astronomy). Large synthesised telescopes like the Cambridge interferometer (also known as the "4-acre array" - the telescope that discovered pulsars and produced the 2C and 3C catalogues) were the first to employ the new technique of aperture synthesis and also boasted up to 16 independent beams separated in declination. In addition, the first radio telescope with an aperture encompassing 1 square kilometre was also realised by Reber via a huge array of dipoles operating at 2 MHz. Though never fully exploited, Reber was certainly ahead of his time in realising the potential of cheaply deploying large-scale apertures at low radio frequencies (see Fig. 1).

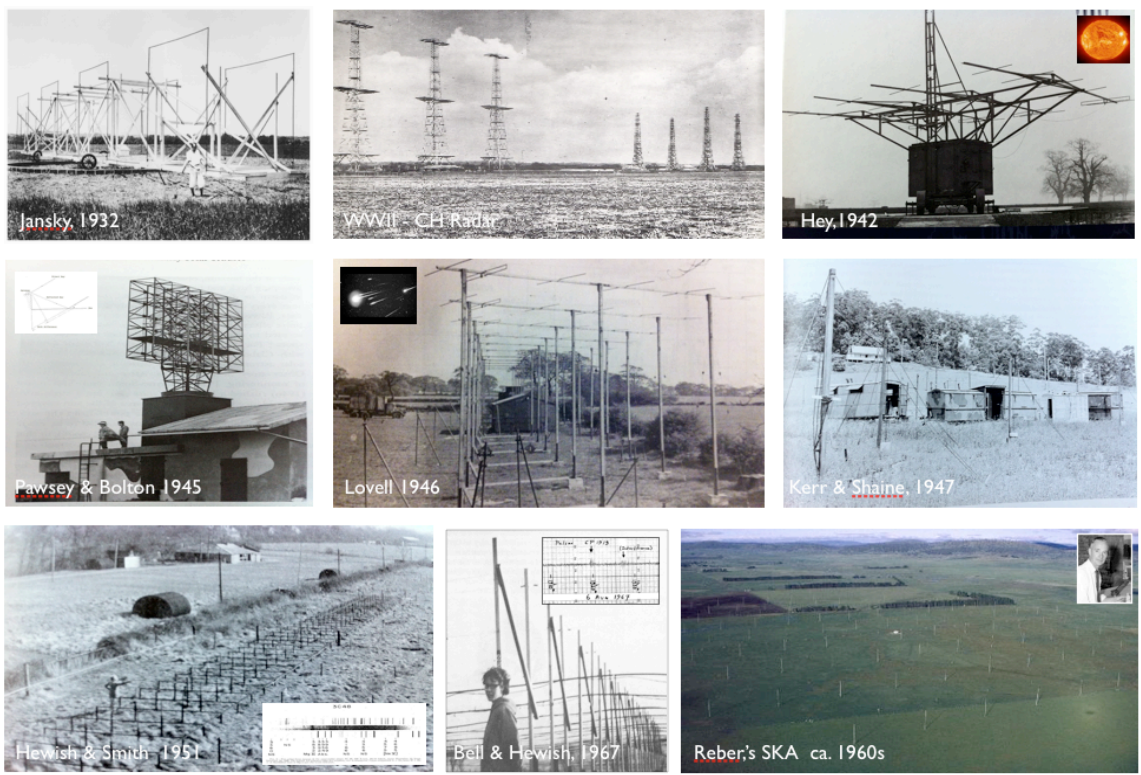

*Figure 1: Some of the developments and aperture array instruments that made such a profound impact on the first few decades of radio astronomy research.*

The foundations of these early radio astronomy discoveries can be traced all the way back to the end of the 19[th] Century. For example, in 1899 the British engineer, Sidney Brown, had connected multiple antennas together in order to achieve directional gain in both transmission and reception – a concept that he patented [2]. In the run up to World War II, countries began to develop active radar systems capable of detecting enemy aircraft at great distances. Sir Robert Watson-Watt, credited with the invention of radar was an irrepressible Scotsman who was responsible for setting up the "Chain Home" (CH) radar system in Britain. Operational just in time for the Battle of Britain (1940), the CH radar masts could detect enemy aircraft at a distance of several hundred kilometres [3]. In the USA, the first electronically steered phased arrays were also under development, first of all for transatlantic communication purposes [4]. At the Rad Lab at MIT, Luis Alvarez built an electronically steered dipole array that became the basis for both the first Approach and Landing System (ALS) and post-WWII early warning systems. After the war, many radar scientists and engineers in Australia, Great Britain & the USA returned to civilian life. On its own, Britain's secret Telecommunications Research Establishment (TRE) furnished several future leaders in the field of radio astronomy such as Bolton, Bowen, Hanbury Brown, Hey, Lovell & Ryle.

## 2. The rise of parabolic dishes

Despite playing a major role in the early stages of radio astronomy, aperture arrays essentially disappeared as a major radio telescope type from the end of the 1960's onwards. Dropping the term "radio telescope" into the Google search engine shows that the field became quickly dominated by parabolic dish systems (see Fig.2). Early post-war examples include the 25-metre Dwingeloo Telescope – the first large, fully steerable antenna for radio astronomy. Until the end of the 20[th] century, parabolic dishes were set to dominate radio astronomy at cm, mm, and sub-mm wavelengths.

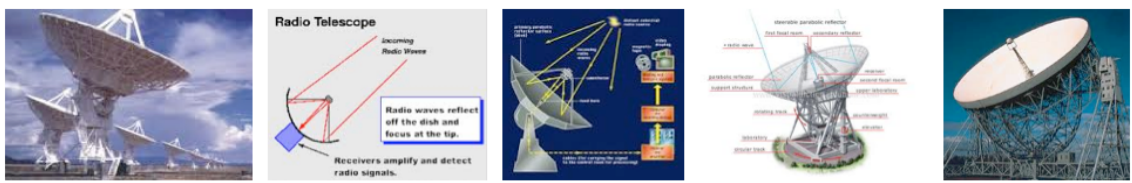

*Figure 2: Typing "radio telescope" into Google images returns these images first. Nothing resembling an Aperture Array appears until the bottom of page 4 of the Google output!*

It is interesting to consider the reasons for this sea change in the field. Certainly, steel parabolas offered the ability to move radio astronomy up to much higher frequencies. This permitted higher resolution measurements to be made which greatly aided the astronomer's task of cross-matching radio sources with existing optical catalogues. In addition, there was a clear need to observe at 21cm – the rest frequency of neutral hydrogen, the most abundant element in the universe. Parabolic shaped antennas also offered much better reliability than aperture arrays, since in those days the latter would contain hundreds of dipoles interconnected by many kilometres of wiring. With a single focal point, parabolas also permitted highly specialised low-noise receivers to be used, increasing their sensitivity by cooling the amplifier systems.

## 2.1 Aperture arrays – a lost technology ?

The disappearance of aperture arrays for the next 40 years (1970-2010), has some parallels with the Antikythera mechanism, and the engineering and science capability it represents (see Seiradakis 2012, these proceedings). Certainly there seems to be no device of comparable complexity built for at least a millennium after the Antikythera mechanism was constructed. Similarly, aperture arrays languished as historical curiosities for many years while large single dishes or arrays of dishes deployed as interferometers dominated the radio astronomy agenda.

## 2.2 Reinventing radio astronomy – the rediscovery of Aperture Arrays

At ASTRON, in the Netherlands, staff began thinking about the feasibility of building a huge radio telescope with a collecting area of about 1 square kilometre ([5] & [6]) – Aperture Arrays began to seem interesting again. Arnold van Ardenne, then Director of ASTRON's R&D labs had recently returned from Ericsson in 1994, and set up an aperture array development line, mostly focused on systems operating at much higher frequencies than before –1-2 GHz (see Fig.3). This led to the production of a series of aperture array tiles: the AAD (Adaptive Antenna Demonstrator), OSMA (One Square Metre Array) [7], THEA (THousand Element Array) and subsequently EMBRACE (Electronic Multi-beam Radio Astronomy ConcEpt) as part of the international FP6 SKADS project [8]. In the slipstream of these developments, the concept of LOFAR also began to gain traction [9,10,11].

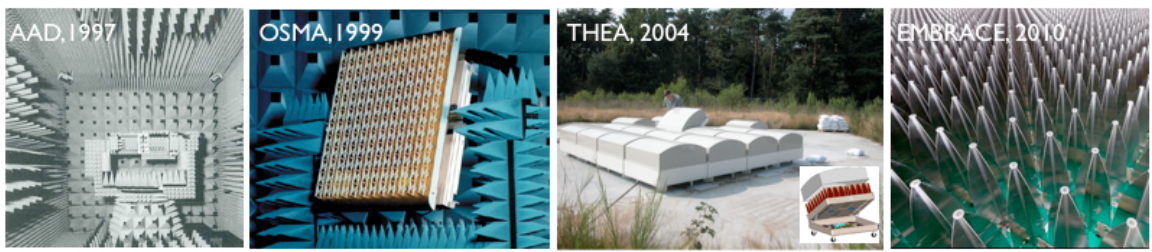

*Figure 3: Four generations of ASTRON's Aperture Array tile development programme - from the early 1990s through to the present day. Images courtesy, A. van Ardenne, J.G. bij de Vaate (ASTRON).*

Technological advances in areas such as signal processing, digital electronics, low-power/high performance super-computing and large capacity data storage systems meant that this new generation of aperture array telescopes would be quite unlike anything that had gone before. The concept of the "Software Telescope" arose, as it was realised that today's electronics permitted many copies of signals to be made at the individual dipole level - independent multiple beams (multiple fields-of-view) could be easily created and rapidly shifted across the sky. High-speed networks, based on new optical fibre technologies also permitted distributed real-time arrays to be constructed, separated by hundreds or even thousands of kilometres. By buffering data at the dipole level via the availability of large, inexpensive RAM modules, retrospective imaging of the entire sky also became possible. In short, the ICT revolution had transformed aperture arrays from an unwieldy, unreliable and forgotten technology of yesteryear, to the basis for a new kind of telescope with the potential to completely transform radio astronomy.

## 3.0 Radio Astronomy Transformed

Modern Aperture Array telescopes [12], offer many advantages over conventional parabolic systems. Figure 4 shows the main elements of the aperture array concept, alongside a traditional parabolic dish. Aperture arrays focus incoming radiation by varying the delay (more usually phase) electronically across the aperture. Dishes focus light via reflection from the parabolic metal surface. An aperture array provides a fully unblocked aperture with an unrestricted view of the entire sky. Composed of simple antennas with commercially available low-noise room-temperature amplifiers, huge collecting areas can be synthesized at relatively low cost. Multiple beams (or multiple fields-of-view) can be rapidly formed and electronically steered across the sky. Retrospective imaging of the sky is possible by buffering raw voltage data at the individual dipole level, and later recombining these data from the entire array with the appropriate delay (i.e. the desired pointing direction). In short, a modern aperture array system provides the ultimate flexibility, reliability and performance with no moving parts involved.

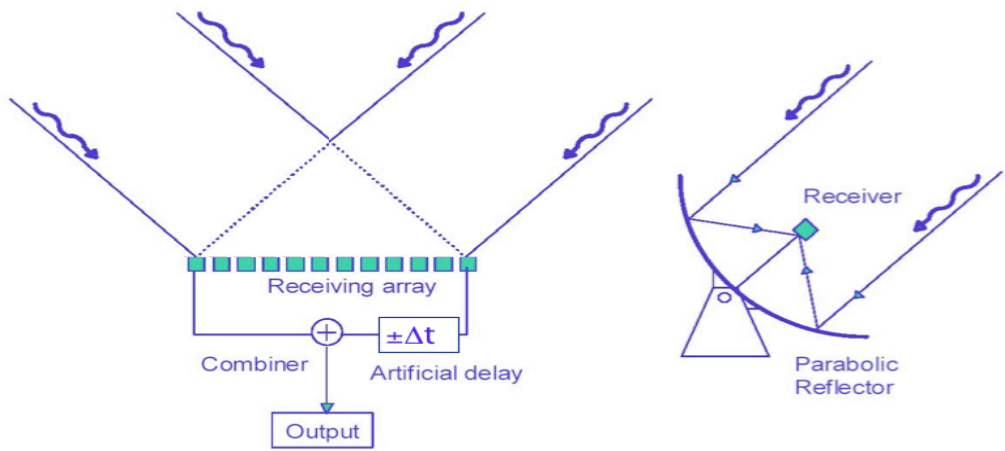

*Figure 4: the main elements of the aperture array concept (left). For comparison, a traditional parabola dish (right). Aperture arrays focus incoming radiation by varying the delay electronically across the aperture. Dishes focus light from a single direction via reflection from the parabolic metal surface.*

### 3.1 Next generation, low-frequency aperture arrays

Since the turn of the 21$^{st}$ century, aperture arrays have reappeared as major elements of the next generation of low-frequency radio telescopes. Fuelled by technological advances and the growing interest to study neutral hydrogen in the early universe, telescopes such as LOFAR [11], the MWA, PAPER and the LWA [13] are currently under construction or being commissioned (see Fig. 5). LOFAR is by far the largest example with the broadest range of observing modes. With baseline lengths extending from the North of the Netherlands, through Germany, Sweden, France & the UK, sub-arcsecond resolution is possible at the highest LOFAR observing frequencies (~ 200 MHz).

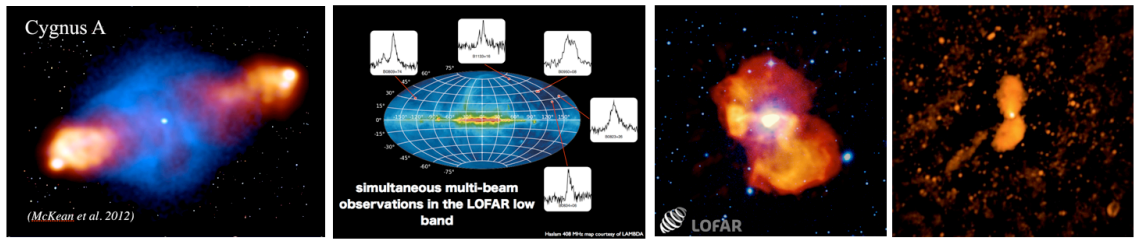

*Figure 5: LOFAR and MWA (right) commissioning results, courtesy Messrs McKean et al., Hessels et al., Gasperin et al. & McKinley et al.*

Figure 6 shows LOFAR and the other low-frequency aperture array telescopes that have emerged over the last few years. The first commissioning results from these telescopes are impressive (see Figure 5), clearly demonstrating the huge scientific potential of aperture array systems.

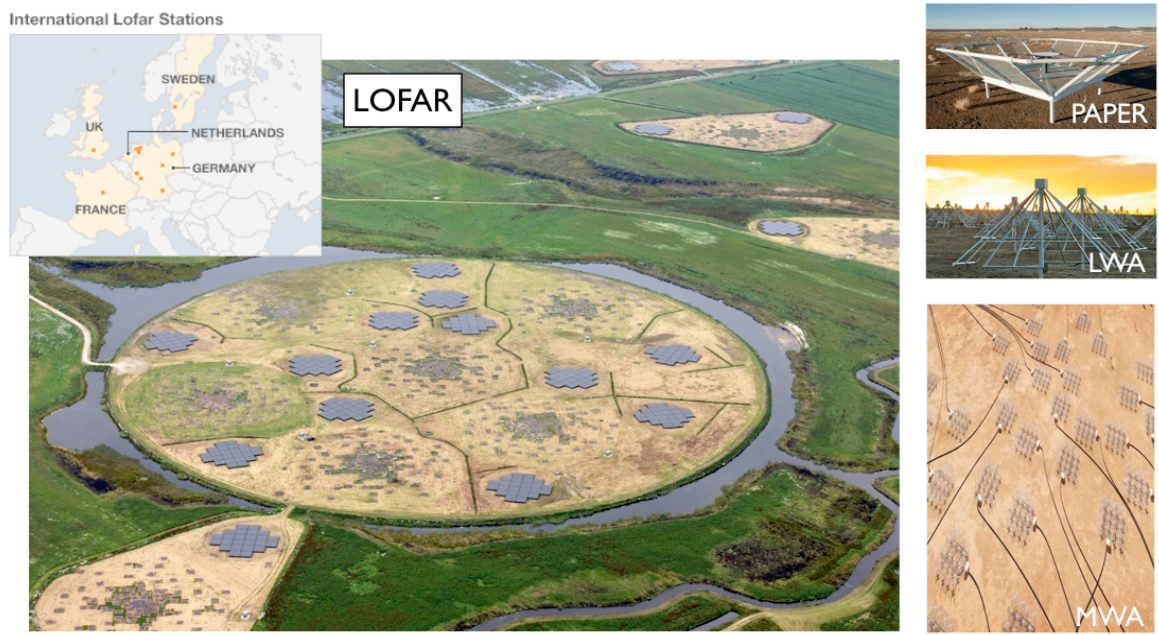

*Figure 6: Radio astronomy reinvented - the 21st century has seen an explosion of aperture array telescopes being constructed – LOFAR (left), the MWA, LWA & PAPER.*

### 3.2 Phased Array Feeds

This revival in aperture array technology development, also led to the realisation that the field of view of traditional parabolic dishes could also be transformed via the introduction of Phased Array Feeds (PAFs). Typically a PAF will extend the field of view of a traditional parabolic dish by a factor of about 30 (see Figure 7). The ability to form telescope beams using the weighted sum of all the receiving antennas also permits very high efficiencies to be achieved via optimal illumination of the dish. Since it is impossible to cool hundreds of amplifiers, this improved efficiency largely compensates for the higher system temperatures of PAFs due to the need to use room temperature amplifiers. However, significant improvements in room-temperature LNAs are expected over the coming years.

These PAFs can be thought of as a 2-dimensional receiver array – something like a "radio camera" with 100s of "pixels" populating the focal plane [9]. In 2006, modified versions of ASTRON's THEA tile (see Fig. 3) were successfully tested first at the focus of CSIRO's Fleurs dish in Sydney and then at one of the WSRT antennas. Based on these early successes, large PAF-based telescope systems like ASKAP and the WSRT-APERTIF are now under construction (see Fig.7).

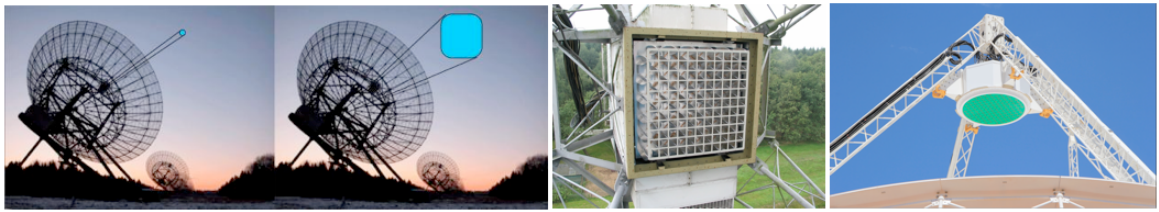

*Figure 7: The huge gain in field-of-view enabled by PAFs. Right-centre: APERTIF installed at the focus of a WSRT telescope. Extreme right: PAFs installed on ASKAP (courtesy CSIRO).*

## 4.0 Aperture Arrays – beyond astronomy

Astronomers are conservative – at least some of them are – and while many have been slow to re-embrace this "old" technology (with a few notable exceptions – see Section 2.1), the rest of the world has adopted aperture array systems in many different guises (see Fig. 8).

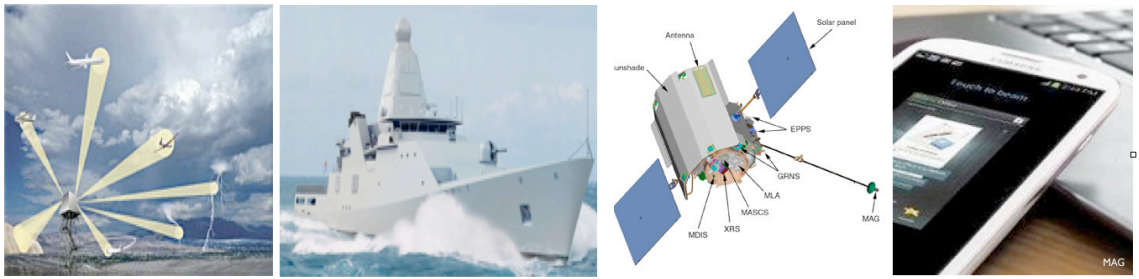

*Figure 8: Aperture arrays (also known elsewhere as smart antennas or adaptive antennas) are now commonly found in many civilian, military and space applications: from left to right - active and passive radar systems (e.g. Thales naval defence systems), deep space communications and in the future mass market mobile/wireless systems.*

The same advantages that aperture arrays offer to radio astronomy (large and multiple field-of-views, rapid electronic steering, reliability, flexibility, cost and performance) are not surprisingly, also considered important in other fields. Aperture arrays also offer some fields other distinct advantages - in the realm of military operations their flat profiles are much easier to seamlessly incorporate into the surface structure of other systems (e.g. vehicles, ships, aircraft fuselage and missile skins) than parabolic dishes. Notably, these systems are typically operating at 1-10 GHz. Military systems are also known to include significant redundancy and "self-healing" features (referred to as "smart skins") in which the systems can be automatically reconfigured after being partially disabled. Aperture Arrays now replace dish systems for both active and passive radar applications (civilian and military systems), weather monitoring, surveillance, navigation and communications (both ground and space). Even in areas that are often reticent to take up new technologies, aperture arrays are present e.g. NASA's Messenger spacecraft, now in orbit around Mercury is the first deep space mission to employ aperture arrays as its main communication system. In the future, the further adoption of so-called steerable "smart antennas" (aperture arrays) in mass-market wireless applications is expected to have a huge impact on the world of mobile and wireless communications [14]. All these new developments will surely have an important impact on the form and design of the SKA, especially on the longer timescales of SKA Phase 2.

## 5.0 The Future of Aperture Arrays and the SKA

Aperture arrays operating at frequencies between 0.1 and 1.5 GHz are naturally one of the major enabling elements that can make the SKA a truly transformational telescope, well into the middle of this century. The study of neutral hydrogen in the local, distant and early universe was the original driver for the SKA, and it still remains the premier scientific case. Phase 1 includes a low-frequency array (SKA1-LO) that naturally builds on pathfinders such as LOFAR and precursors such as the MWA in Western Australia. Phase 2 is envisaged to include a major Dense Aperture Array (DAA) component, operating at frequencies of ~ 0.5-1.5 GHz to be sited in South Africa. There is a clear need to continue to advance DAA tile developments and to construct a significant precursor telescope before embarking on the construction of a much larger DAA in Phase 2. Currently known as "EMMA" [15], we propose the construction of a 16-station SKA-2 precursor array becoming operational well before the end of the decade (see Fig. 9). This ambitious timescale should be considered in the context of the current SKA timeline.

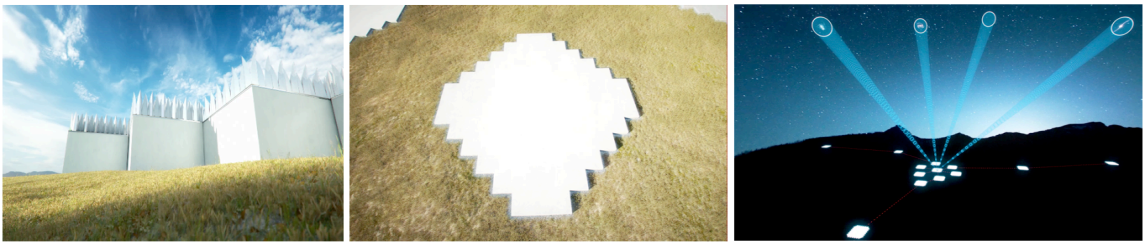

*Figure 9: EMMA – a SKA-2 Dense Aperture Arrays precursor telescope.*

Like the Antikythera mechanism itself (see these proceedings), the SKA must be an instrument that will amaze future generations to come! Building a SKA that is simply the "VLA on steroids" is simply not good enough – we have the ability to do much, *much* better. The present generation of astronomers, and the generations to come, deserve nothing less.